\documentclass[12pt]{article}
\usepackage[dvips]{graphicx}  

\begin{document}

\newcommand{\rum}{\rule{0.5pt}{0pt}}

\newcommand{\rub}{\rule{1pt}{0pt}}

\newcommand{\rim}{\rule{0.3pt}{0pt}}

\newcommand{\numtimes}{\mbox{\raisebox{1.5pt}{${\scriptscriptstyle \times}$}}}

\renewcommand{\refname}{References}


\begin{center}

{\Large\bf  Dynamical 3-Space:  Supernovae  and the Hubble \\
\vspace{3mm}
Expansion -  Older Universe and End of  Dark Energy  \rule{0pt}{13pt}}\par

\bigskip

Reginald T. Cahill \\ 

{\small\it School of Chemistry, Physics and Earth Sciences, Flinders University,

Adelaide 5001, Australia\rule{0pt}{13pt}}\\

\raisebox{-1pt}{\footnotesize E-mail: Reg.Cahill@flinders.edu.au}\par

\bigskip\smallskip

{\small\parbox{11cm}{%

We apply the new dynamics of 3-space to cosmology by deriving a Hubble expansion solution.  This dynamics involves two constants; $G$ and $\alpha$ - the fine structure constant. This solution gives an excellent parameter-free fit to the recent supernova and gamma-ray burst data without the need for `dark energy' or `dark matter'.  The data and theory together imply an older age for the universe of some 14.7Gyrs.  Various problems such as fine tuning, the event horizon problem etc are now resolved.  A brief review  discusses the origin of the 3-space dynamics  and how that dynamics explained the bore hole anomaly, spiral galaxy flat rotation speeds, the masses of black holes in spherical galaxies,   gravitational light bending and lensing, all without invoking `dark matter' or `dark energy'.  These developments imply that a new  understanding of the universe is now available.

\rule[0pt]{0pt}{0pt}}}\bigskip

\end{center}

\section{Introduction}

There are theoretical claims based on  observations of Type Ia supernova (SNe Ia)\cite{S1,S2} that the universe expansion is accelerating. The cause of this acceleration has been attributed to an undetected `dark energy'.   Here the dynamical theory of 3-space is applied to Hubble expansion dynamics, with the result that the supernova and gamma-ray burst data is well fitted without an acceleration effect and without the need to introduce any notion of `dark energy'.  So, like `dark matter', `dark energy' is an unnecessary and spurious notion.  A brief review is included showing how the 3-space dynamics arises and how that dynamics explained the bore hole anomaly, spiral galaxy flat rotation speeds, the masses of black holes in spherical galaxies,   gravitational light bending and lensing, all without invoking `dark matter' or `dark energy'.  These developments imply that a new  understanding of the universe is now available. 

\section{The Physics of 3-Space - a Review}

\subsection{ 3-Space Dynamics}
At a deeper level an information-theoretic approach to modelling reality, {\it Process Physics} \cite{Book, Review},  leads to an emergent structured `space'  which is 3-dimensional and dynamic, but where the 3-dimensionality is only approximate, in that if we ignore non-trivial topological aspects of the space, then it may be embedded in a 3-dimensional  geometrical manifold.  Here the space is a real existent discrete but fractal network of relationships or connectivities,  but the embedding space is purely a mathematical way of characterising the 3-dimensionality of the network.  This is illustrated in Fig.1. Embedding the network in the embedding space is very arbitrary; we could equally well rotate the embedding or use an embedding that has the network translated or translating.  These general requirements  then dictate the minimal dynamics for the actual network, at a phenomenological level.  To see this we assume  at a coarse grained level that the dynamical patterns within the network may be described by a velocity field ${\bf v}({\bf r},t)$, where ${\bf r}$ is the location of a small region in the network according to some arbitrary embedding.  The 3-space velocity field has been observed in at least 8 experiments  [3-15].
For simplicity we assume here that the global topology of the network   is not significant for the local dynamics, and so we embed in an $E^3$, although a generalisation to an embedding in $S^3$ is straightforward and might be relevant to cosmology.  The minimal dynamics is then obtained by  writing down the lowest-order zero-rank tensors, of dimension $1/t^2 $, that are invariant under translation and rotation, giving.

\begin{figure}[t]
\vspace{3.2mm}\,\hspace{15mm}\parbox{60mm}{\includegraphics[width=60mm]{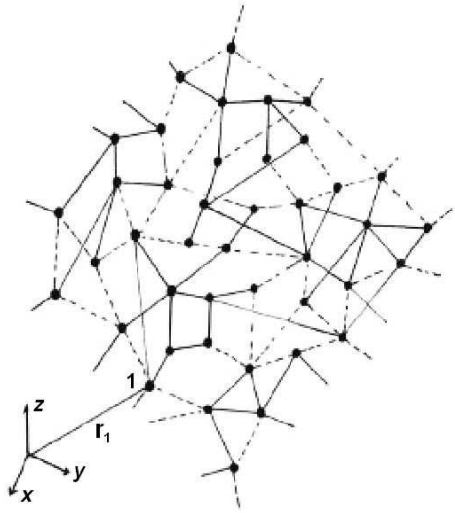}}
\parbox{70mm}{\caption{\small{  This is an iconic graph\-ic\-al representation of how a dynamical  network has its inherent approximate 3-dimen\-sion\-al\-ity displayed by an embedding in a math\-em\-at\-ical   space such as an $E^3$ or an $S^3$.  This space is not real; it is purely a mathematical artifact. Nevertheless this em\-bedd\-abi\-li\-ty helps de\-term\-ine the minimal dyn\-am\-ics for the network, as in (\ref{eqn:E1}).  At a deep\-er \,level \,the \,network \,is \,a \,quantum foam system \cite{Book}. The dyn\-am\-ical space is not an ether model, as the em\-bedd\-ing space does not exist. The dyn\-am\-ical space is not an ether model, as the em\-bedd\-ing space does not exist.}}}
\label{fig:Embedd}
\end{figure}

\begin{figure}[t]
\vspace{3.2mm}\,\hspace{5mm}\parbox{70mm}{\includegraphics[width=70mm]{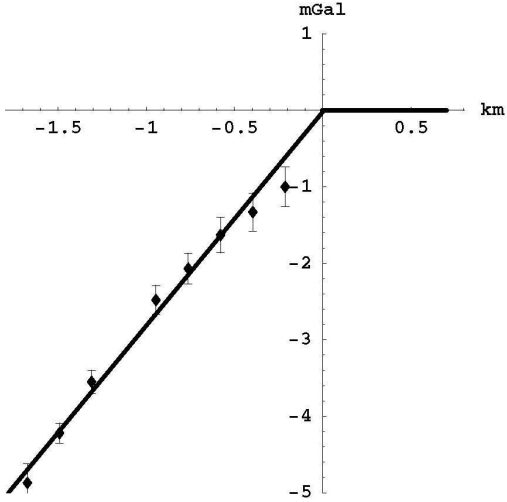}}\,
\parbox{70mm}{\caption{\small{  The data shows the gravity residuals for the Greenland Ice Shelf \cite{Ander89} Airy measurements of
the $g(r)$  profile,  defined as $\Delta g(r) = g_{Newton}-g_{observed}$, and measured in mGal (1mGal $ =10^{-3}$ cm/s$^2$)
and   plotted against depth in km. The borehole effect is that Newtonian
gravity and the new theory differ only beneath the surface, provided that the measured above surface gravity gradient 
is used in  both theories.  This then gives the horizontal line above the surface. Using (\ref{eqn:E6}) we obtain
$\alpha^{-1}=137.9 \pm  5$ from fitting the slope of the data, as shown. The non-linearity  in the data arises from
modelling corrections for the gravity effects of the   irregular sub ice-shelf rock  topography. The dyn\-am\-ical space is not an ether model, as the em\-bedd\-ing space does not exist.}}}
\label{fig:Greenland}
\end{figure}

\begin{figure}[t]
\vspace{3.2mm}\,\hspace{10mm}\parbox{70mm}{\includegraphics[width=70mm,scale=0.2]{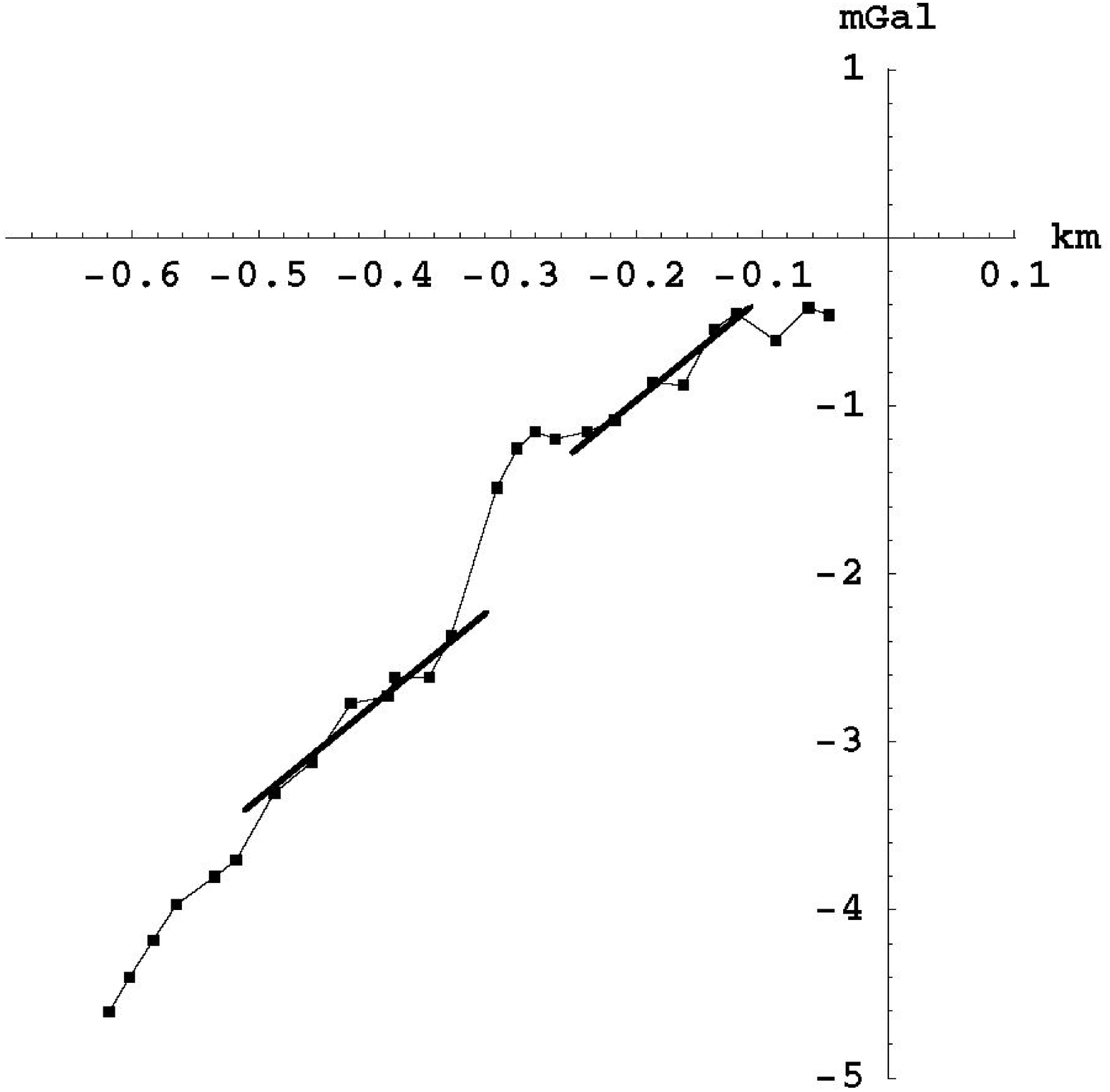}}
\parbox{70mm} {\includegraphics[width=70mm,scale=0.2]{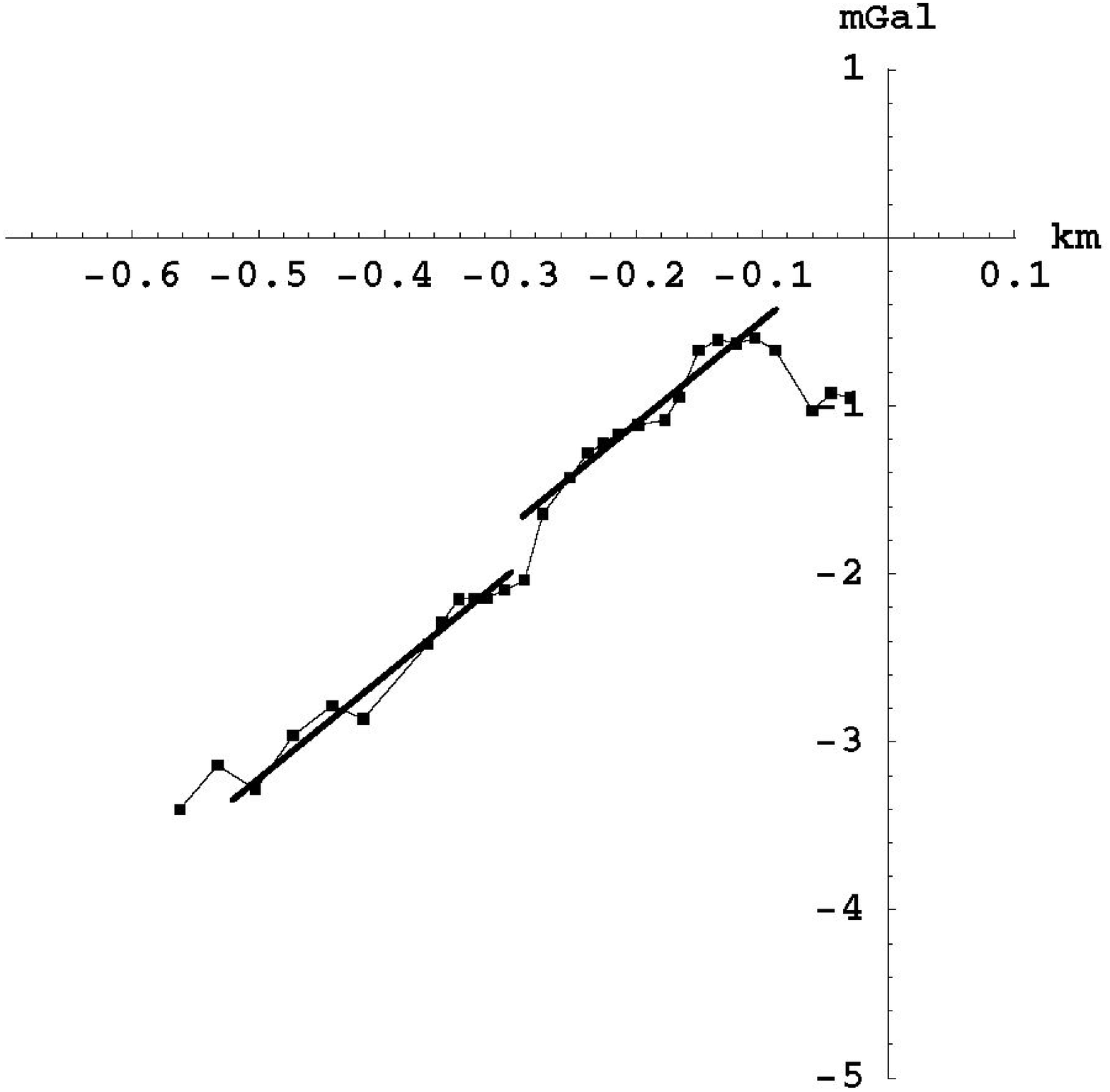}}
{\caption{\small{  Gravity residuals from two of the Nevada bore hole experiments \cite{Thomas90} that give a best fit of $\alpha^{-1}=136.8\pm 3$ on using (\ref{eqn:E6}). Some layering of the rock is evident.}}}
\label{fig:Nevada}
\end{figure}

\begin{equation}
\nabla.\left(\frac{\partial {\bf v} }{\partial t}+({\bf v}.{\bf \nabla}){\bf v}\right)
+\frac{\alpha}{8}(tr D)^2 +\frac{\beta}{8}tr(D^2)=
-4\pi G\rho
\label{eqn:E1}\end{equation}
\begin{equation}D_{ij}=\frac{1}{2}\left(\frac{\partial v_i}{\partial x_j}+
\frac{\partial v_j}{\partial x_i}\right)
\label{eqn:E1b}\end{equation}
where $\rho({\bf r},t)$ is the effective matter density. The embedding space coordinates provide a coordinate system or frame of reference that is convenient to describing the velocity field, but which is not real.  
In Process Physics  quantum matter  are topological defects in the network, but here it is sufficient to give a simple description in terms of an  effective density. 

We see that there are only four possible terms, and so we need at most three possible constants to describe the dynamics of space: $G, \alpha$ and $\beta$. $G$ turns out  to be Newton's gravitational constant, and describes the rate of non-conservative flow of space into matter.  To determine the values of $\alpha$ and $\beta$ we must, at this stage, turn to experimental data.  
However most experimental data involving the dynamics of space is observed by detecting the so-called gravitational  acceleration of matter, although increasingly light bending is giving new information.  Now the acceleration ${\bf a}$ of the dynamical patterns in space is given by the Euler or convective expression
\begin{equation}
{\bf a}({\bf r},t)= \lim_{\Delta t \rightarrow 0}\frac{{\bf v}({\bf r}+{\bf v}({\bf r},t)\Delta t,t+\Delta
t)-{\bf v}({\bf r},t)}{\Delta t} 
=\frac{\partial {\bf v}}{\partial t}+({\bf v}.\nabla ){\bf v}
\label{eqn:E3}\end{equation} 
and this appears in one of the terms in (\ref{eqn:E1}). As shown in \cite{Schrod} and discussed later herein the acceleration  ${\bf g}$ of quantum matter is identical to this acceleration, apart from vorticity and relativistic effects, and so the gravitational acceleration of matter is also given by (\ref{eqn:E3}).

Outside of a spherically symmetric distribution of matter,  of total mass $M$, we find that one solution of (\ref{eqn:E1}) is the velocity in-flow field  given by
\begin{equation}
{\bf v}({\bf r})=-\hat{{\bf r}}\sqrt{\frac{2GM(1+\frac{\alpha}{2}+..)}{r}}
\label{eqn:E4}\end{equation}
but only when $\beta=-\alpha$,  for only then is the acceleration of matter, from (\ref{eqn:E3}), induced by this in-flow of the form
\begin{equation}
{\bf g}({\bf r})=-\hat{{\bf r}}\frac{GM(1+\frac{\alpha}{2}+..)}{r^2}
\label{eqn:E5}\end{equation}
 which  is Newton's Inverse Square Law of 1687 \cite{Newton}, but with an effective  mass $M(1+\frac{\alpha}{2}+..)$ that is different from the actual mass $M$.  So the success of Newton's law in the solar system informs us that  $\beta=-\alpha$ in (\ref{eqn:E1}). But we also see modifications coming from the 
$\alpha$-dependent terms.

In general because (\ref{eqn:E1}) is a scalar equation it is only applicable for vorticity-free flows $\nabla\times{\bf v}={\bf 0}$, for then we can write ${\bf v}=\nabla u$, and then (\ref{eqn:E1}) can always be solved to determine the time evolution of  $u({\bf r},t)$ given an initial form at some time  $t_0$.
The $\alpha$-dependent term in (\ref{eqn:E1})  (with now $\beta=-\alpha$) and the matter acceleration effect, now also given by (\ref{eqn:E3}),   permits   (\ref{eqn:E1})   to be written in the form
\begin{equation}
\nabla.{\bf g}=-4\pi G\rho-4\pi G \rho_{DM},
\label{eqn:E7}\end{equation}
where
\begin{equation}
\rho_{DM}({\bf r},t)\equiv\frac{\alpha}{32\pi G}( (tr D)^2-tr(D^2)),  
\label{eqn:E7b}\end{equation}
which  is an effective matter density that would be required to mimic the
 $\alpha$-dependent spatial self-interaction dynamics.
 Then (\ref{eqn:E7}) is the differential form for Newton's law of gravity but with an additional non-matter effective matter density.  So we label this as $\rho_{DM}$ even though no matter is involved \cite{alpha,DM}. This effect has been shown to explain the so-called `dark matter' effect in spiral galaxies, bore hole $g$ anomalies, and the systematics of galactic black hole masses.  
 
 The spatial dynamics  is non-local.  Historically this was first noticed by Newton who called it action-at-a-distance. To see this we can write  (\ref{eqn:E1}) as an integro-differential equation
 \begin{equation}
 \frac{\partial {\bf v}}{\partial t}=-\nabla\left(\frac{{\bf v}^2}{2}\right)+G\!\!\int d^3r^\prime
 \frac{\rho_{DM}({\bf r}^\prime, t)+\rho({\bf r}^\prime, t)}{|{\bf r}-{\bf r^\prime}|^3}({\bf r}-{\bf r^\prime})
 \label{eqn:E8}\end{equation}

 This shows a high degree of non-locality and non-linearity, and in particular that the behaviour of both $\rho_{DM}$ and $\rho$ manifest at a distance irrespective of the dynamics of the intervening space. This non-local behaviour is analogous to that in quantum systems and may offer a resolution to the horizon problem.

 \subsection{ Bore Hole Anomaly}
 
 A recent discovery \cite{alpha, DM} has been that experimental data from the bore hole $g$ anomaly has revealed that $\alpha$ is the fine structure constant, to within experimental errors: $\alpha=e^2/\hbar c \approx 1/137.04$. This anomaly is that $g$ does not decrease as rapidly as predicted by Newtonian gravity or GR as we descend down a bore hole.  
 Consider the case where we have a spherically symmetric matter distribution at rest, on average with respect to distant space, and that the in-flow is time-independent and radially symmetric.  Then (\ref{eqn:E1})  can now be written in the form,  with $v^\prime=dv(r)/dr$, 
 \begin{equation}
2\frac{vv^\prime}{r} +(v^\prime)^2 + vv^{\prime\prime} =-4\pi G\rho(r)-4\pi G \rho_{DM}(v(r)), 
\label{eqn:Eradial}
\end{equation} 
where
 \begin{equation}
\rho_{DM}(r)= \frac{\alpha}{8\pi G}\left(\frac{v^2}{2r^2}+ \frac{vv^\prime}{r}\right).
\label{eqn:E10}\end{equation}
The dynamics in (\ref{eqn:Eradial}) and (\ref{eqn:E10}) gives
 the anomaly to be
 \begin{equation}
 \Delta g=2\pi\alpha G \rho d +O(\alpha^2)
 \label{eqn:E6}\end{equation}
where $d$ is the depth and $\rho$ is the density, being that of glacial ice in the case of the Greenland Ice Shelf experiments \cite{Ander89}, or that of rock in the Nevada test site experiment \cite{Thomas90}. Clearly (\ref{eqn:E6}) permits the value of $\alpha$ to be determined from the data, giving  $\alpha=1/ (137.9 \pm 5)$ from the Greenland data, and  $\alpha=1/(136.8\pm 3)$ from the Nevada data; see Figs.\ref{fig:Greenland} and \ref{fig:Nevada}. 

\subsection{ Black Holes}

\begin{figure}[t]
\hspace{0mm}\includegraphics[scale=0.5]{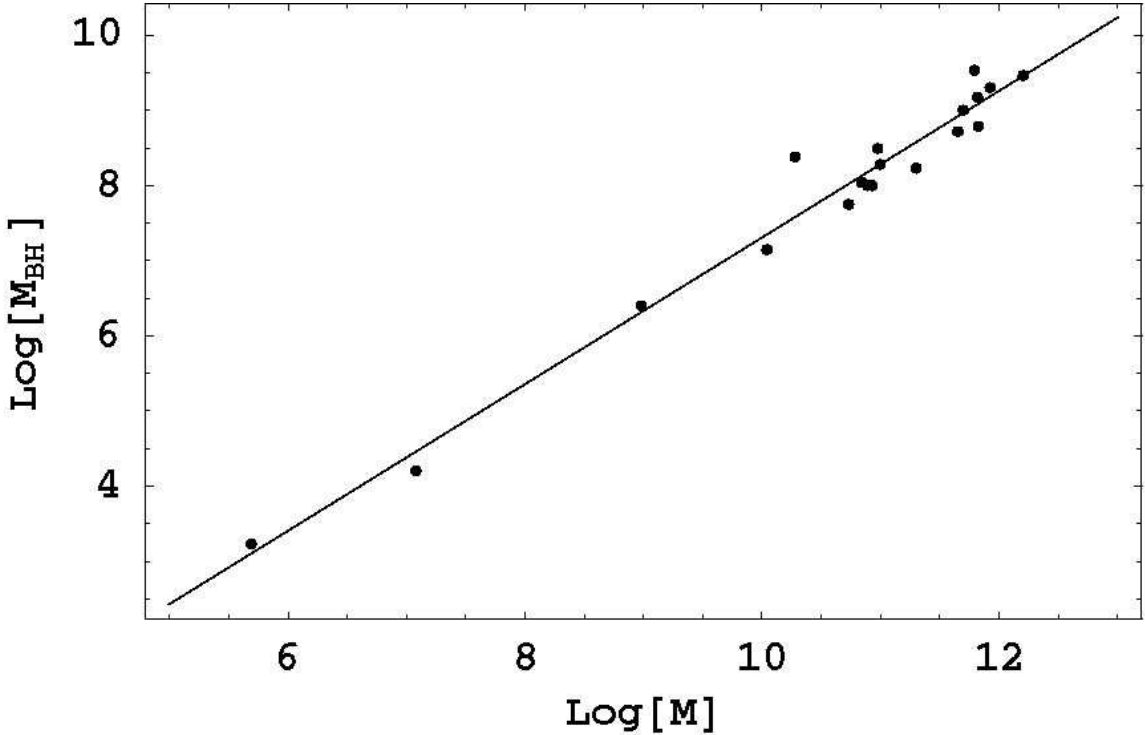}
\vspace{-4mm}
\caption{\small{The data shows $\mbox{Log}_{10}[M_{BH}]$ for the black hole masses $M_{BH}$  for
a variety of spherical matter systems with masses $M$, plotted against 
$\mbox{Log}_{10}[M]$, in solar masses $M_0$.  The straight line is the prediction from (\ref{eqn:bhmasses}) with $\alpha=1/137$. See \cite{newBH} for references to the data. 
  \label{fig:blackholes}}}
\end{figure}

Eqn.(\ref{eqn:Eradial}) with $\rho=0$ has  exact analytic `black hole' solutions.  For minimal black holes induced by a spherically symmetric distribution of matter  we find by iterating (\ref{eqn:Eradial}) and then from  (\ref{eqn:E10}) that the total effective black hole  mass is
\begin{equation}
M_{BH}=M_{DM} = 4\pi\int_0^\infty r^2\rho_{DM}(r)dr = \frac{\alpha}{2}M+O(\alpha^2)
\label{eqn:bhmasses}\end{equation}
This solution is applicable to the black holes at the centre of spherical star systems, where we identify $M_{DM}$ as $M_{BH}$.    So far black holes in 19  spherical star systems have been detected and together their masses are plotted in 
Fig.\ref{fig:blackholes} and compared with (\ref{eqn:bhmasses}) \cite{galaxies,newBH}.

\subsection{Spiral Galaxy Rotation Anomaly}

\begin{figure}[t]
\hspace{20mm}\includegraphics[scale=1.3]{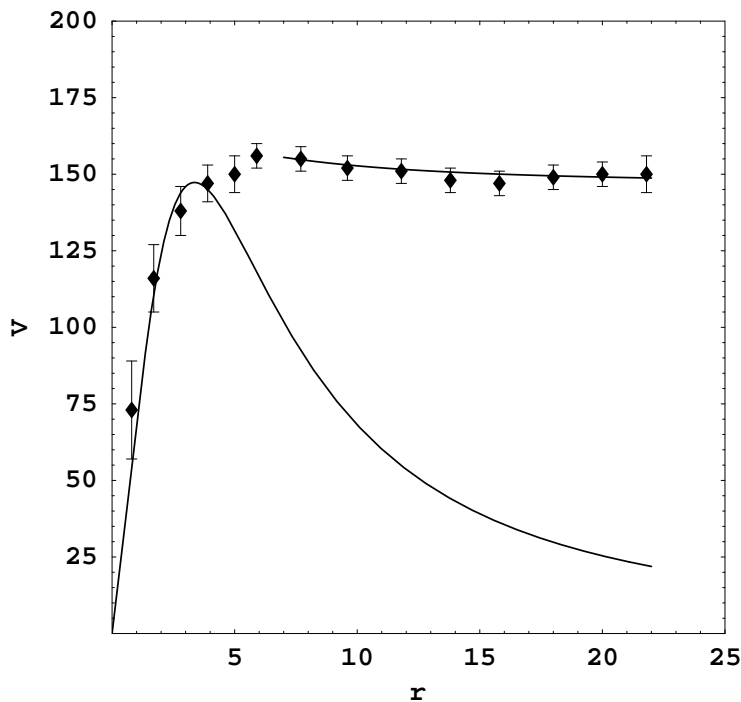}
\vspace{-4mm}
\caption{\small {Data shows the non-Keplerian rotation-speed curve $v_O$ for the spiral galaxy NGC 3198 in km/s plotted
against radius in kpc/h. Lower curve is the rotation curve from the Newtonian theory  for an
exponential disk, which decreases asymptotically like $1/\sqrt{r}$. The upper curve shows the asymptotic form from
(\ref{eqn:vorbital}), with the decrease determined by the small value of $\alpha$.  This asymptotic form is caused by
the primordial black holes at the centres of spiral galaxies, and which play a critical role in their formation. The
spiral structure is caused by the rapid in-fall towards these primordial black holes.}
\label{fig:NGC3198}}\end{figure}

Eqn  (\ref{eqn:Eradial}) gives also a direct explanation for the spiral galaxy rotation anomaly.    For a non-spherical system numerical solutions of (\ref{eqn:E1}) are required, but sufficiently far from the centre we find an exact non-perturbative two-parameter class of analytic solutions
\begin{equation}
v(r) = K\left(\frac{1}{r}+\frac{1}{R_s}\left(\frac{R_s}{r}  \right)^{\displaystyle{\frac{\alpha}{2}}}  \right)^{1/2}
\label{eqn:vexactb}\end{equation}
where $K$ and $R_s$ are arbitrary constants in the $\rho=0$ region, but whose values are determined by matching to the solution in the matter region. Here $R_s$ characterises the length scale of the non-perturbative part of this expression,  and $K$ depends on $\alpha$, $G$ and details of the matter distribution.   From (\ref{eqn:E5})  and (\ref{eqn:vexactb}) we obtain a replacement for  the Newtonian  `inverse square law' ,
\begin{equation}
g(r)=\frac{K^2}{2} \left( \frac{1}{r^2}+\frac{\alpha}{2rR_s}\left(\frac{R_s}{r}\right)
^{\displaystyle{\frac{\alpha}{2}}} 
\right),
\label{eqn:gNewl}\end{equation}
in the asymptotic limit.     The centripetal acceleration  relation for circular orbits 
$v_O(r)=\sqrt{rg(r)}$  gives  a `universal rotation-speed curve'
\begin{equation}
v_O(r)=\frac{K}{2} \left( \frac{1}{r}+\frac{\alpha}{2R_s}\left(\frac{R_s}{r}\right)
^{\displaystyle{\frac{\alpha}{2}}} 
\right)^{1/2}
\label{eqn:vorbital}\end{equation}
 Because of the $\alpha$ dependent part this rotation-velocity curve  falls off extremely slowly with $r$, as is indeed observed for spiral galaxies. An example is shown in Fig.\ref{fig:NGC3198}. It was the inability of the  Newtonian  and Einsteinian gravity theories to explain these observations that led to the  notion of `dark matter'.

\subsection{Generalised  Maxwell Equations}

 We must   generalise the Maxwell equations so that the electric and magnetic  fields are excitations of the dynamical space. 
 \begin{equation}
\displaystyle{ \nabla \times {\bf E}=-\mu\left(\frac{\partial {\bf H}}{\partial t}+{\bf v.\nabla H}\right)},
 \mbox{\ \ \ }\displaystyle{\nabla.{\bf E}={\bf 0}},
\end{equation}
 \begin{equation} \displaystyle{ \nabla \times {\bf H}=\epsilon\left(\frac{\partial {\bf E}}{\partial t}+{\bf v.\nabla E}\right)} ,
\mbox{\ \ \  }\displaystyle{\nabla.{\bf H}={\bf 0}}\label{eqn:E18}\end{equation}
which was first suggested by Hertz in 1890  \cite{Hertz}, but with ${\bf v}$ being a constant vector field. As easily determined  the speed of EM radiation is now $c=1/\sqrt{\mu\epsilon}$ with respect to the 3-space, and in general not with respect to the observer if the observer is moving through space, as experiment has indicated again and again  [3-15].
In particular the in-flow in (\ref{eqn:E4}) causes a refraction effect of light passing close to the sun, with the angle of deflection given by
\begin{equation}
\delta=2\frac{v^2}{c^2}=\frac{4GM(1+\frac{\alpha}{2}+..)}{c^2d}
\label{eqn:E19}\end{equation}
where $v$ is the in-flow speed at distance $d$  and $d$ is the impact parameter, here the radius of the sun. Hence the  observed deflection of $8.4\times10^{-6}$ radians is actually a measure of the in-flow speed at the sun's surface, and that gives $v=615$km/s.   These generalised Maxwell equations also predict gravitational lensing produced by the large in-flows that are the  new `black holes' in galaxies.  \subsection{Generalised  Schr\"{o}dinger Equation}

A  generalisation of the  Schr\"{o}dinger equation  is \cite{Schrod} is
\begin{equation}
i\hbar\frac{\partial  \psi({\bf r},t)}{\partial t}=H(t)\psi({\bf r},t),
\label{eqn:equiv7}\end{equation}
where the free-fall hamiltonian is uniquely
\begin{equation}
H(t)=-i\hbar\left({\bf
v}.\nabla+\frac{1}{2}\nabla.{\bf v}\right)-\frac{\hbar^2}{2m}\nabla^2
\label{eqn:equiv8}\end{equation}
This follows from  the wave function being attached to the dynamical space, and not to the embedding space, and that $H(t)$ be hermitian. We can compute the acceleration of a localised wave packet  according to
\begin{equation}{\bf g}\equiv\frac{d^2}{dt^2}\left(\psi(t),{\bf r}\psi(t)\right)  
=\frac{\partial{\bf v}}{\partial t}+({\bf v}.\nabla){\bf v}+
(\nabla\times{\bf v})\times{\bf v}_R
\label{eqn:E11}\end{equation}
where ${\bf v}_R={\bf v}_0-{\bf v}$  is the velocity of the wave packet relative to the local space, as ${\bf v}_0$ is  the velocity relative to the embedding space. Apart from the vorticity term which causes rotation of the wave packet\footnote{This explains the Lense-Thirring effect, and such vorticity  is being detected by the Gravity Probe B satellite gyroscope experiment\cite{GPB}.} we see, as promised, that this matter acceleration is equal to that of the space itself, as in (\ref{eqn:E3}). This is the first derivation of the phenomenon of gravity from a deeper theory: gravity is a quantum effect - namely the refraction of quantum waves by the internal differential motion of the substructure  patterns to space itself. Note that the equivalence principle has now been explained, as this `gravitational' acceleration is independent of the mass $m$ of the quantum system. 

\subsection{Generalised  Dirac Equation}
An analogous generalisation of the Dirac equation is also necessary giving the coupling of the spinor to the actual dynamical 
3-space, and again not to the embedding space as has been the case up until now, 
\begin{equation}
i\hbar\frac{\partial \psi}{\partial t}=-i\hbar\left(  c{\vec{ \alpha.}}\nabla + {\bf
v}.\nabla+\frac{1}{2}\nabla.{\bf v}  \right)\psi+\beta m c^2\psi
\label{eqn:12}\end{equation}
where $\vec{\alpha}$ and $\beta$ are the usual Dirac matrices. Repeating the analysis in (\ref{eqn:E11}) for the space-induced acceleration we obtain
\begin{equation}\label{eqn:E12}
{\bf g}=\displaystyle{\frac{\partial {\bf v}}{\partial t}}+({\bf v}.{\bf \nabla}){\bf
v}+({\bf \nabla}\times{\bf v})\times{\bf v}_R-\frac{{\bf
v}_R}{1-\displaystyle{\frac{{\bf v}_R^2}{c^2}}}
\frac{1}{2}\frac{d}{dt}\left(\frac{{\bf v}_R^2}{c^2}\right)
\label{eqn:E13a}\end{equation}
which generalises  (\ref{eqn:E11}) by having a term which limits the speed of the wave packet relative to space to be $<\!c$. This equation specifies the trajectory of a spinor wave packet in the dynamical 3-space.

\subsection{Deriving the Spacetime and Geodesic Formalism\label{section:spacetime}}

 We find that (\ref{eqn:E12}) may be obtained by extremising the time-dilated elapsed time 
\begin{equation}
\tau[{\bf r}_0]=\int dt \left(1-\frac{{\bf v}_R^2}{c^2}\right)^{1/2}
\label{eqn:E13}\end{equation}  
with respect to the particle trajectory ${\bf r}_0(t)$ \cite{Book}. This happens because of the Fermat least-time effect for waves: only along the minimal time trajectory do the quantum waves  remain in phase under small variations of the path. This again emphasises  that gravity is a quantum effect.   We now introduce an effective  spacetime mathematical construct according to the metric
\begin{equation}
ds^2=dt^2 -(d{\bf r}-{\bf v}({\bf r},t)dt)^2/c^2 
=g_{\mu\nu}dx^{\mu}dx^\nu
\label{eqn:E14}\end{equation}
Then according to this metric the elapsed time in (\ref{eqn:E13}) is
\begin{equation}
\tau=\int dt\sqrt{g_{\mu\nu}\frac{dx^{\mu}}{dt}\frac{dx^{\nu}}{dt}},
\label{eqn:E14b}\end{equation}
and the minimisation of  (\ref{eqn:E14b}) leads to the geodesics of the spacetime, which are thus equivalent to the trajectories from (\ref{eqn:E13}), namely (\ref{eqn:E13a}).
Hence by coupling the Dirac spinor dynamics to the 3-space  we derive the geodesic formalism of General Relativity as a quantum effect, but without reference to the Hilbert-Einstein equations for the induced metric.  Indeed in general the metric of  this induced spacetime will not satisfy  these equations as the dynamical space involves the $\alpha$-dependent  dynamics, and $\alpha$ is missing from GR.   Nevertheless it may be shown \cite{Book} that in the limit $\alpha\rightarrow 0$  the induced metric in (\ref {eqn:E14}), with ${\bf v}$ from   (\ref {eqn:E1}), satisfies the Hilbert-Einstein equations so long as we use relativistic corrections for the matter density on the RHS of (\ref {eqn:E1}). This means that  (\ref {eqn:E1}) is consistent with for example the binary pulsar data - the relativistic aspects being associated with the matter effects upon space and the relativistic effects of the matter in motion through the dynamical 3-space.  The agreement of GR with the pulsar data is implying that the $\alpha$-dependent effects are small in this case, unlike in black holes and spiral galaxies.

\section{Supernova and Gamma-Ray Burst Data}

The supernovae and gamma-ray bursts provide standard candles that enable observation  of the expansion of the universe. 
 The supernova data set used herein and shown in Figs.\ref{fig:SN1} and \ref{fig:SN2} is available at \cite{data set}.    Quoting from  \cite{data set}  we note that Davis {\it et al.} \cite{Davis}  combined several data sets by taking  the ESSENCE data set from Table 9 of Wood--Vassey {\it et al.}  (2007) \cite{WV}, using only the supernova that passed the light-curve-fit quality criteria. They took the HST data from Table 6 of Riess {\it et al.} (2007) \cite{Riess}, using only the supernovae classified as gold.
To put these data sets on the same Hubble diagram  Davis {\it et al.} used 36 local supernovae that are in common between these two data sets. When discarding supernovae with $z<0.0233$ (due to larger uncertainties in the peculiar velocities) they found an offset of $0.037 \pm 0.021$ magnitude between the data sets, which they then corrected for by subtracting this constant from the HST data set. The dispersion in this offset was also accounted for in the uncertainties.
The HST data set had an additional 0.08 magnitude added to the distance modulus errors to allow for the intrinsic dispersion of the supernova luminosities. The value used by Wood--Vassey {\it et al.}  (2007) \cite{WV} was instead 0.10 mag. Davis {\it  et al.}  adjusted for this difference by putting the Gold supernovae on the same scale as the ESSENCE supernovae. Finally, they also added the dispersion of 0.021 magnitude introduced by the simple offset described above to the errors of the 30 supernovae in the HST data set. The final supernova data base for  the distance modulus $\mu_{obs}(z)$ is shown in Figs.\ref{fig:SN1} and \ref{fig:SN2}.  The gamma-ray burst (GRB) data is from Schaefer \cite{GRB}.

\section{Expanding 3-Space - the Hubble Solution}

Suppose that  we have a radially symmetric density $\rho(r,t)$ and that we look for a radially symmetric time-dependent flow ${\bf v}({\bf r},t) =v(r,t)\hat{\bf r}$ from (\ref{eqn:E1}) (with $\beta=-\alpha$).  Then $v(r,t)$ satisfies the equation,  with $v^\prime=\displaystyle{\frac{\partial v(r,t)}{\partial r}}$,
\begin{equation}
\frac{\partial}{\partial t}\left( \displaystyle{\frac{2v}{r}}+v^\prime\right)+vv^{\prime\prime}+2\frac{vv^{\prime}}{r}+ (v^\prime)^2+\frac{\alpha}{4}\left(\frac{v^2}{r^2} +\frac{2vv^\prime}{r}\right)
=- 4\pi G \rho(r,t)  \label{eqn:radialflow}\end{equation}

Consider first the zero energy case $\rho=0$. Then we have a Hubble  solution $v(r,t)=H(t)r$, a centreless flow, determined by
\begin{equation}{\dot H}+\left(1+\frac{\alpha}{4}\right)H^2=0
\end{equation}
with ${\dot H}=\displaystyle{\frac{dH}{dt}}$.  We also introduce in the usual manner the scale factor $R(t)$ according to $H(t)=\displaystyle{\frac{1}{R}\frac{dR}{dt}}$. We then obtain
the solution
\begin{equation}
H(t)=\frac{1}{(1+\frac{\alpha}{4})t}=H_0\frac{t_0}{t}; \mbox{\ \  }  R(t)=R_0\left(\frac{t}{t_0} \right)^{4/(4+\alpha)}
\label{eqn:spacexp}\end{equation}
where $H_0=H(t_0)$ and $R_0=R(t_0)$. 
We can write the  Hubble function $H(t)$ in terms of $R(t)$ via the inverse function $t(R)$, i.e. $H(t(R))$ and finally as $H(z)$, where the redshift observed now, $t_0$, relative to the wavelengths at time $t$, is  $z=R_0/R-1$. Then we obtain
\begin{eqnarray}
H(z)={H_0}(1+z)^{1+\alpha/4}
\label{eqn:H2a}\end{eqnarray}

We need to determine the distance travelled by the light from a supernova before detection. Using a choice of coordinate system with $r=0$ at the location of a supernova the 
speed of light relative to this embedding space frame is $c+v(r(t),t)$, i.e. $c$ wrt the space itself, as noted above,  where $r(t)$ is the distance from the source. Then the distance travelled by the light at time $t$ after emission at time $t_1$ is determined implicitly by
\begin{equation}
r(t)=\int_{t_1}^t dt^\prime(c+v(r(t^\prime), t^\prime),
\label{eqn:distance1}\end{equation}
which has the solution on using $v(r,t)=H(t)r$
\begin{equation}
r(t)=c R(t)\int_{t_1}^t \frac{dt^\prime}{R(t^\prime)}
\label{eqn:distance2}\end{equation}
Expressed in terms of the observable redshift $z$ this gives
\begin{equation}
r(z)=c (1+z)\int_{0}^z \frac{dz^\prime}{H(z^\prime)}
\label{eqn:distance3}\end{equation}
 The effective dimensionless distance is given by
 \begin{equation}
d(z)=(1+z)\int_0^z \frac{H_0 dz^\prime}{H(z^\prime)}
\label{eqn:H1a}\end{equation}
 and the theory distance modulus is then defined by
\begin{equation}
\mu_{th}(z)=5\log_{10}(d(z))+m
\label{eqn:H1b}\end{equation}
Because all the selected supernova have the same absolute magnitude, $m$ is a constant whose value is determined by fitting the low $z$ data.

Using the  Hubble expansion (\ref{eqn:H2a}) in (\ref{eqn:H1a}) and (\ref{eqn:H1b}) we obtain the middle curves (red) in Figs.\ref{fig:SN1} and  the \ref{fig:SN2}, yielding an excellent agreement with the supernovae and GRB data. Note that because $\alpha/4$ is so small it actually has negligible effect on these plots.  Hence the dynamical 3-space gives an immediate account of the universe expansion data, and does not require the introduction of  a cosmological constant or `dark energy', but which will be nevertheless discussed next.

When the energy density is not zero we need to take account of the dependence of $\rho(r,t)$ on the scale factor of the universe. In the usual manner we thus write
\begin{equation}
\rho(r,t)=\frac{\rho_{m}}{R(t)^3}+\frac{\rho_{r}}{R(t)^4}+\Lambda
\end{equation}
for matter, EM radiation and the cosmological constant or `dark energy' $\Lambda$, respectively, where the matter and radiation is approximated by a spatially uniform (i.e independent of $r$)  equivalent matter density. We argue here that $\Lambda$ - the dark energy density, like dark matter, is an unnecessary concept.
Then (\ref{eqn:radialflow}) becomes for $R(t)$
\begin{eqnarray}
\frac{\ddot  R}{R}+\frac{\alpha}{4}\frac{{\dot R}^2}{R^2}
=-\frac{4\pi G}{3}\left(\frac{\rho_{m}}{R^3}+\frac{\rho_{r}}{R^4}+\Lambda \right)
\end{eqnarray}giving
\begin{equation}
{\dot R}^2=\frac{8\pi G}{3}\left(\frac{\rho_{m}}{R}+\frac{\rho_{r}}{R^2}+\Lambda R^2\right)-\frac{\alpha}{2}\int\frac{{\dot R}^2}{R}dR
\label{eqn:R2}\end{equation}
 In terms of ${\dot R}^2$ this has the solution
\begin{equation}{\dot R}^2\!=\!\frac{8\pi G}{3}\!\!\left(\!\frac{\rho_{m}}{(1-\frac{\alpha}{2})R}\!+\!\frac{\rho_{r}}{(1-\frac{\alpha}{4})R^2}\!+\!\frac{\Lambda R^2}{(1+\frac{\alpha}{4})}\!+\!b R^{-\alpha/2}\!\right)
\label{eqn:R3}\end{equation}
which is easily checked by substitution into (\ref{eqn:R2}), and where $b$ is an arbitrary integration constant. Finally we obtain from  (\ref{eqn:R3})
\begin{equation}
t(R)=\int^R_{R_0}\frac{dR}{\sqrt{\displaystyle{\frac{8\pi G}{3}}\left(\displaystyle{\frac{\rho_{m}}{R}+\frac{\rho_{r}}{R^2}}+\Lambda R^2+b R^{-\alpha/2}\right)}}
\label{eqn:R4}\end{equation} 
where now we have re-scaled parameters $\rho_m\rightarrow\rho_m/(1-\frac{\alpha}{2}), \rho_r\rightarrow\rho_r/(1-\frac{\alpha}{4})$ and $\Lambda\rightarrow\Lambda/(1+\frac{\alpha}{4})$. When $\rho_m=\rho_r=\Lambda=0$, (\ref{eqn:R4})
reproduces the expansion in (\ref{eqn:spacexp}), and so the density terms in (\ref{eqn:R3}) give the modifications to  the dominant purely spatial expansion, which we have noted above  already gives an excellent account of the data.

From (\ref{eqn:R4})  we then obtain
\begin{equation}
H(z)^2={H_0}^2(\Omega_m(1+z)^3+\Omega_r(1+z)^4 +\Omega_\Lambda+\Omega_s(1+z)^{2+\alpha/2})
\label{eqn:H2}\end{equation}
with
\begin{equation}
\Omega_m+\Omega_r+\Omega_\Lambda+\Omega_s=1.
\end{equation}

\begin{figure}
\hspace{-3mm}{\includegraphics[scale=0.5]{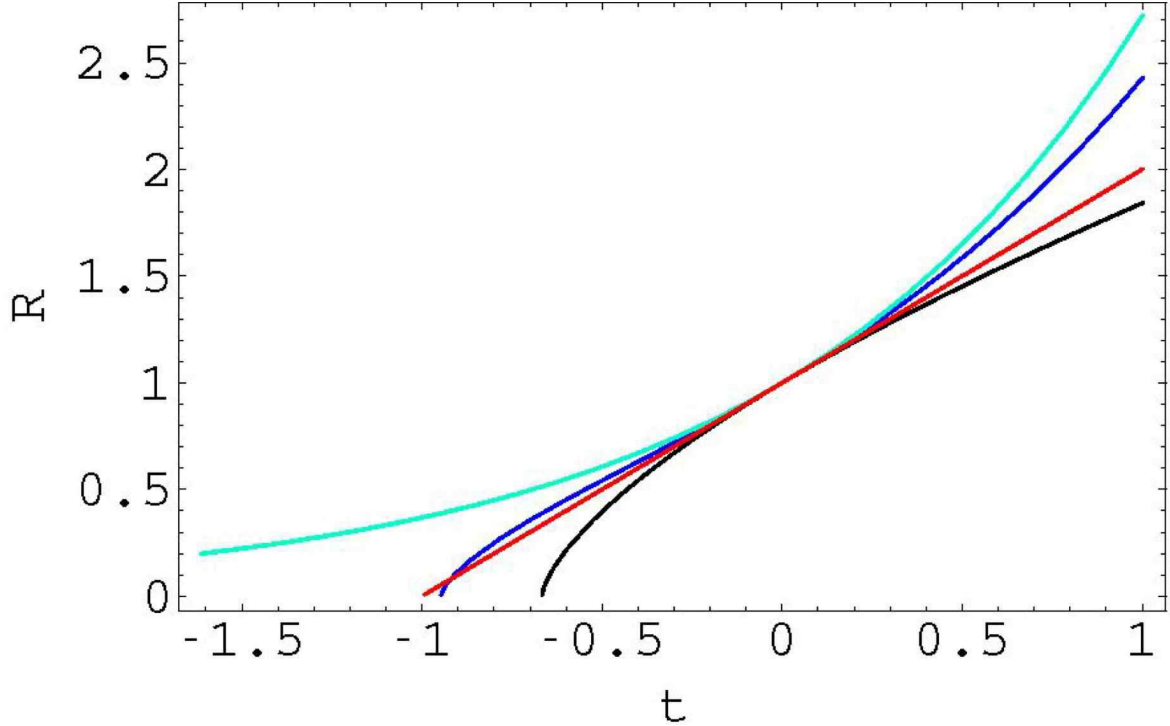}}\,
{\vspace{-4mm}\caption{\small{  Plot of the scale factor $R(t)$ vs $t$, with $t=0$ being `now' with $R(0)=1$, for the four cases discussed in the text, and corresponding to the plots in  Figs.\ref{fig:SN1} and \ref{fig:SN2}: (i) the upper  curve (green) is the `dark energy' only case, resulting in an exponential acceleration at all times, (ii) the bottom curve (black) is the matter only prediction, (iii) the 2nd highest curve (to the right of $t=0$) is the best-fit  `dark energy' plus matter case (blue) showing a past deceleration and future exponential acceleration effect. The straight line plot  (red)  is the dynamical 3-space prediction showing a slightly  older universe compared to case (iii). We see that the best-fit `dark energy' - matter curve essentially converges on the dynamical 3-space result. All plots have the same slope at $t=0$, i.e. the same value of $H_0$. If the age of the universe is inferred to be some 14Gyrs for case (iii) then the age of the universe is changed to some 14.7Gyr for case (iv). 
 }\label{fig:Rtplot}}}
\end{figure}

Using the Hubble function  (\ref{eqn:H2}) in (\ref{eqn:H1a}) and (\ref{eqn:H1b}) we obtain  the plots in  Figs.\ref{fig:SN1} and \ref{fig:SN2} for four cases: (i) $\Omega_m=0,  \Omega_r=0, \Omega_\Lambda=1, \Omega_s=0$,  i.e a pure `dark energy' driven expansion, (ii)  $\Omega_m=1,  \Omega_r=0, \Omega_\Lambda=0, \Omega_s=0$  showing that a matter only expansion is not a good account of the data, (iii) from a least squares fit with $\Omega_s=0$ we find $\Omega_m=0.28,  \Omega_r=0, \Omega_\Lambda=0.68$  which led to the suggestion that `dark energy' effect was needed to fix the poor fit from (ii), and finally  (iv) $\Omega_m=0,   \Omega_r=0,  \Omega_\Lambda=0,  \Omega_s=1$, as noted above, that the spatial expansion dynamics alone gives a good account of the data. Of course the EM radiation term $\Omega_r$ is non-zero but small and determines the expansion during the baryogenesis initial phase, as does the spatial dynamics expansion term because of the $\alpha$ dependence. If the age of the universe is inferred to be some 14Gyrs for case (iii) then, as seen in Fig.\ref{fig:Rtplot}, the age of the universe is changed to some 14.7Gyr for case (iv). We see that the one-parameter  best-fit `dark energy'-matter curve, the $\Lambda CDM$ model,  essentially converges on the no-parameter dynamical 3-space result.

\begin{figure}
\vspace{0mm}
\hspace{-5mm}\includegraphics[scale=0.5]{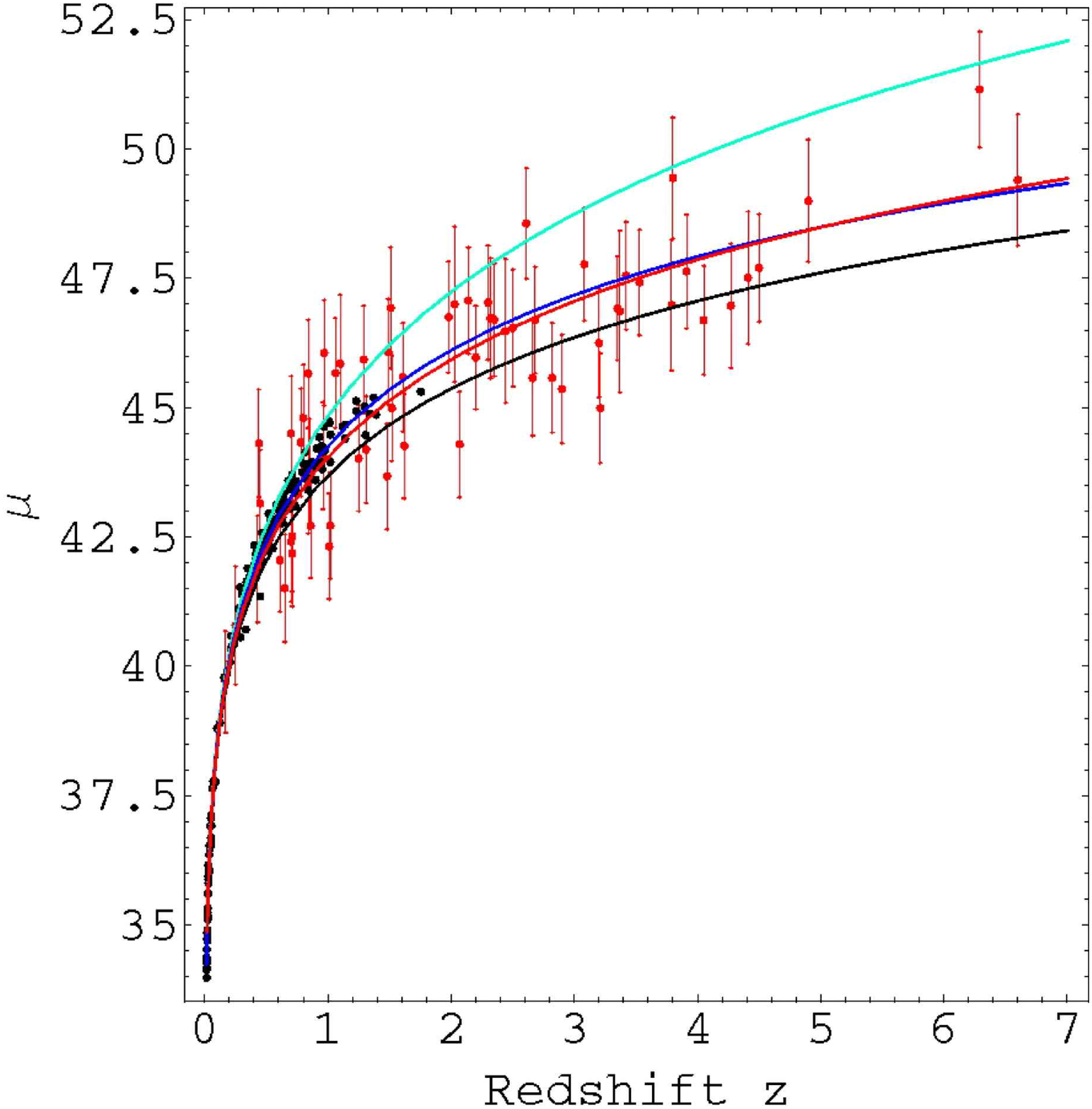}
\vspace{-5mm}\caption{\small{ Hubble diagram showing the combined supernovae data from Davis {\it et al.} \cite{Davis} using several data sets from   Riess {\it et al.} (2007)\cite{Riess} and Wood-Vassey {\it et al.}  (2007)\cite{WV} (dots without error bars for clarity - see Fig.\ref{fig:SN2} for error bars) and the Gamma-Ray Bursts data (with error bars) from Schaefer \cite{GRB}.  Upper curve (green)  is `dark energy' only $\Omega_\lambda=1$, lower curve  (black) is matter only $\Omega_m=1$. Two middle curves show best-fit of `dark energy'-matter (blue) and dynamical 3-space prediction (red), and are essentially indistinguishable.  However the theories make very different predictions for the future and for the age of the universe. We see that the best-fit `dark energy'-matter curve essentially converges on the dynamical 3-space prediction.}
\label{fig:SN1}}\end{figure}

\begin{figure}
\vspace{3.0mm}
\hspace{15mm}{\includegraphics[scale=1.2]{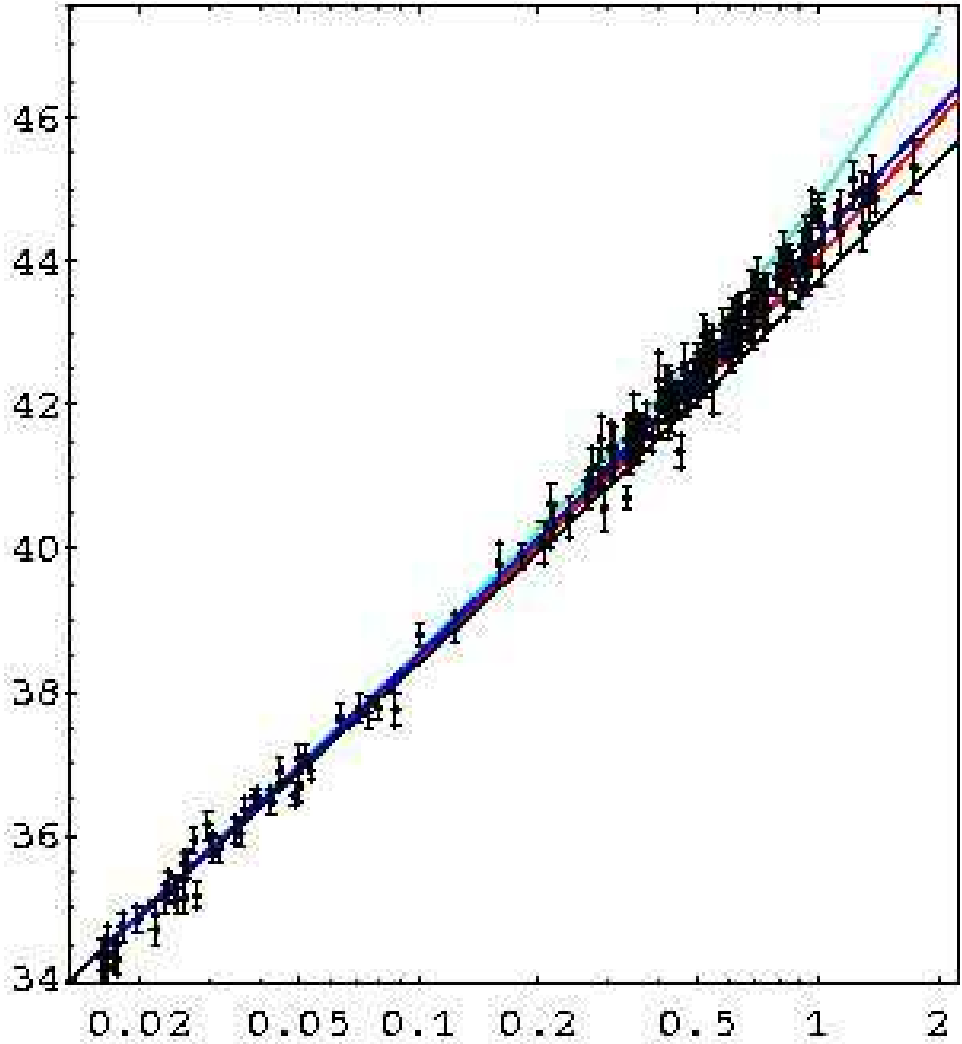}}\,
{\caption{\small{  Hubble diagram as in Fig.\ref{fig:SN1} but plotted logarithmically  to reveal details for  $z<2$, and without GRB data.  Upper curve (green) is `dark energy' only $\Omega_\Lambda=1$. Next curve (blue) is  best fit of `dark energy'-matter. Lowest curve (black) is matter only $\Omega_m=1$. 2nd lowest curve (red) is  dynamical 3-space prediction. }\label{fig:SN2}}}
\end{figure}

\section{Effective Spacetime Metric}
The induced effective spacetime metric in (\ref{eqn:E14}) is for the Hubble expansion
\begin{equation}
ds^2=g_{\mu\nu}dx^\mu dx^\nu=dt^2-(d{\bf r}-H(t){\bf r})dt)^2/c^2 
\label{PGmetric}\end{equation}
 The occurrence of $c$ has nothing to do with the dynamics of the 3-space - it is related to the geodesics of relativistic quantum matter, as noted above. Changing  variables  ${\bf r}\rightarrow R(t){\bf r}$
we obtain
\begin{equation}
ds^2=g_{\mu\nu}dx^\mu dx^\nu=dt^2-R(t)^2d{\bf r}^2/c^2 
\label{Hubblemetric}\end{equation}
which is the usual   Friedman-Robertson-Walker (FRW) metric in the case of a flat spatial section. However when solving for $R(t)$ using the Hilbert-Einstein GR equations
the $\Omega_s$ term (with $\alpha\rightarrow 0$) is usually only present when the spatial curvature is 
non-zero.  So some problem appears to be present in the usual GR analysis of the FRW metric. However above we see that that term arises in fact even when the embedding space is flat.

\section{Conclusions}

We have briefly reviewed the extensive evidence for a dynamical 3-space, with the minimal dynamical equation now known and confirmed by numerous experimental and observational data.  As well we have shown that this equation has a Hubble expanding 3-space solution that in a parameter-free manner manifestly fits the recent supernova and gamma-ray burst data, and in doing so reveals that `dark energy', like `dark matter', is an unnecessary notion.  The Hubble solution leads to a uniformly expanding universe, and so  without acceleration: the claimed acceleration is merely an artifact related to the  unnecessary `dark energy' notion.  This result gives an older age for the universe of some 14.7Gyr, and resolves as well various problems such as the fine turning problem, the horizon problem and other difficulties in the current modelling of the universe.


\begin{thebibliography}{99}

\bibitem{S1} Riess A.G. {\it et al.} {\it Astron. J.} {\bf 116}, 1009, 1998.
\bibitem{S2} Perlmutter S. {\it et al.} {\it Astrophys. J.} {\bf 517}, 565, 1999.
\bibitem{Book} Cahill  R.T. {\it Process Physics: From Information Theory to Quantum Space
       and Matter},  Nova Science Pub., New York, 2005.
       \bibitem{Review} Cahill, R.T. {\it Dynamical 3-Space: A Review},  arXiv:0705.4146.
\bibitem{MM}  Michelson A.A. and  Morley E.W. {\it Philos. Mag.} S.5  24 No.151, 449-463, 1887.
\bibitem{MMCK} Cahill R.T. and Kitto K. {\it Michelson-Morley Experiments Revisited}, {\it Apeiron}, {\bf 10}(2),104-117, 2003.
 \bibitem{AMGE}  Cahill R.T. {\it Absolute Motion and Gravitational Effects}, {\it Apeiron},   {\bf 11}(1), 53-111, 2004.
  \bibitem{MMC}  Cahill  R.T. {\it The Michelson and Morley 1887 Experiment
and the Discovery of Absolute Motion},   {\it Progress in Physics},  {\bf 3}, 25-29, 2005.
\bibitem{Miller}   Miller D.C. {\it Rev. Mod. Phys.},  {\bf 5}, 203-242, 1933.
\bibitem{C5}   Illingworth K.K. {\it  Phys. Rev.} 3,  692-696, 1927.
\bibitem{C6}   Joos G. {\it  Ann. d. Physik} [5] 7,  385, 1930.
\bibitem{C7}   Jaseja T.S. {\it et al.} {\it  Phys. Rev.} A 133, 1221, 1964.
\bibitem{Torr}  Torr D.G. and Kolen P. in  {\it Precision Measurements and Fundamental Constants},  Taylor, B.N. and  Phillips, W.D.  eds.{\it  Natl. Bur. Stand. (U.S.), Spec. Pub.}, 617,  675, 1984.
\bibitem{DeWitte}  Cahill R.T. {\it The Roland DeWitte 1991 Experiment}, {\it Progress in Physics}, {\bf 3}, 60-65, 2006.
\bibitem{Coax} Cahill R.T. {\it  A New Light-Speed Anisotropy Experiment:  Absolute Motion and Gravitational Waves Detected}, {\it Progress in Physics}, {\bf 4}, 73-92, 2006. \bibitem{Schrod} Cahill R.T. {\it  Dynamical  Fractal  3-Space and the Generalised Schr\"{o}dinger  
Equation: Equivalence Principle and  Vorticity Effects},   {\it Progress in Physics},  {\bf 1}, 27-34, 2006.
\bibitem{Newton}  Newton I.  {\it Philosophiae Naturalis Principia Mathematica}, 1687.
\bibitem{alpha}  Cahill R.T.   {\it Gravity, `Dark Matter' and the Fine Structure Constant}, {\it Apeiron}, {\bf
12}(2), 144-177, 2005.
 \bibitem{DM}   Cahill R.T.   {\it  `Dark Matter' as a Quantum Foam In-flow Effect}, in {\it
Trends in Dark Matter Research},  96-140,  ed. J. Val Blain , Nova Science Pub., New York, 2005.    
\bibitem{Ander89}  Ander M.E. {\it et al.} {\it Test of Newton's inverse-square law in the Greenland Ice
Cap}, {\it Phys. Rev. Lett.},  {\bf 62},  985-988, 1989.
\bibitem{Thomas90} Thomas J. and Vogel P. {\it Testing the inverse-square law of gravity in boreholes at the
Nevada test site}, {\it Phys. Rev. Lett.},   {\bf 65},  1173-1176, 1990.
\bibitem{galaxies}  Cahill   R.T. {\it Black Holes in Elliptical and Spiral Galaxies and in 
Globular Clusters}, {\it Progress in Physics}, {\bf 3}, 51-56, 2005.
\bibitem{newBH} Cahill  R.T. {\it Black Holes and Quantum Theory: The Fine Structure Constant Connection},  {\it Progress in Physics}, {\bf 4}, 44-50, 2006.
\bibitem{Hertz}  Hertz H.  {\it On the Fundamental Equations of Electro-Magnetics for Bodies in Motion}, {\it Wiedemann's Ann.}  {\bf 41}, 369, 1890;  {\it Electric Waves, Collection of Scientific Papers,}  {\it Dover Pub.}, New  York, 1962.
\bibitem{GPB} Cahill  R.T. {\it Novel Gravity Probe B Frame-Dragging Effect},  {\it Progress in Physics}, {\bf 3}, 30-33, 2005.
\bibitem{data set} http://dark.dark-cosmology.dk/ $\sim$tamarad/SN/
\bibitem{Davis} Davis  T., Mortsell E., Sollerman J. and  ESSENCE,  {\it  Scrutinizing Exotic Cosmological Models Using ESSENCE Supernovae Data Combined with Other Cosmological Probes},  astro-ph/0701510, 2007.
\bibitem{Riess} Riess A.G.  {\it et al.},  {\it  New Hubble Space Telescope Discoveries of Type Ia Supernovae at  $z > 1$: Narrowing Constraints on the Early Behavior of Dark Energy}, astro-ph/0611572, 2007.
\bibitem{WV} Wood-Vassey  W.M. {\it et al.},  {\it Observational Constraints on the Nature of the Dark Energy: First Cosmological Results from the ESSENCE Supernovae Survey},  astro-ph/0701041, 2007.
\bibitem{GRB} Schaefer B.E. {\it  The Hubble Diagram to Redshift $ >$  6 from 69 Gamma-Ray Bursts},
{\it Ap. J.} {\bf 660}, 16-46, 2007.
\end{thebibliography}
\end{document}